\documentclass[12pt,a4paper]{article}
\makeatletter
\def\eqnarray{%
\stepcounter{equation}%
\let\@currentlabel=\theequation
\global\@eqnswtrue
\global\@eqcnt\z@
\tabskip\@centering
\let\\=\@eqncr
$$\halign to \displaywidth\bgroup\@eqnsel\hskip\@centering
$\displaystyle\tabskip\z@{##}$&\global\@eqcnt\@ne
\hfil$\displaystyle{{}##{}}$\hfil
&\global\@eqcnt\tw@$\displaystyle\tabskip\z@{##}$\hfil
\tabskip\@centering&\llap{##}\tabskip\z@\cr}
\makeatother
\newcommand{\ket}[1]{{\vert{#1}\rangle}}
\newcommand{\bra}[1]{{\langle{#1}\vert}}
\newcommand{\kett}[2]{{\vert{#1,#2}\rangle}}
\newcommand{\braa}[2]{{\langle{#1,#2}\vert}}
\newcommand{\calh}{{\cal H}}

\newcommand{\fukuso}{{\mathbf C}}
\newcommand{\futon}{{\bf N}}

\newcommand{\wzetta}{{\vert w\vert}}
\newcommand{\zetta}{{\vert z\vert}}

\newcommand{\chizeta}{{\vert\chi\vert}}

\newcommand{\kappazetta}{{\vert\kappa\vert}}

\begin{document}

\title{\sl Basic Properties of Coherent and Generalized Coherent 
           Operators Revisited}
\author{
  Kazuyuki FUJII
  \thanks{E-mail address : fujii@math.yokohama-cu.ac.jp }\\
  Department of Mathematical Sciences\\
  Yokohama City University\\
  Yokohama, 236-0027\\
  Japan
  }
\date{}
\maketitle\thispagestyle{empty}
%
%
%
%
\begin{abstract}
  In this letter we make a brief review of some basic properties 
  (the matrix elements, the trace, the Glauber formula) of 
  coherent operators and study the corresponding ones for generalized 
  coherent operators based on Lie algebra su(1,1). We also propose some 
  problems.
\end{abstract}

\newpage

%
%
%
%

\section{Introduction}

Coherent states or generalized coherent states play an important role in 
quantum physics, in particular, quantum optics, see \cite{KS} and its 
references. They also play an important one in mathematical physics. See 
the book \cite{AP}. For example, they are very useful in performing 
stationary phase approximations to path integral, \cite{FKSF1}, 
\cite{FKSF2}, \cite{FKS}. 

Coherent operators which produce coherent states are very useful because 
they are unitary. They are also easy to handle. Why are they so handy ? 
The basic reason is probably that they are subject to the elementary 
Baker-Campbell-Hausdorff formula. 
Many basic properties of them have been investigated,  see \cite{KS}. 
In this letter we are particularly interested in the following three 
ones : the matrix elements, the trace and the Glauber's formula. 
See \cite{AP} or \cite{GLM} more recently . 

Generalized coherent operators which produce generalized coherent states 
are also useful. But the corresponding properties have not been 
investigated as far as the author knows. The one of reasons is that they 
are not so easy to handle. Of course we have the disentangling formula 
corresponding to the elementary Baker-Campbell-Hausdorff formula, but 
they are not handy too.

In this letter we investigate the three properties above for generalized 
coherent operators based on Lie algebra $su(1,1)$. But we don't treat 
generalized coherent operators based on Lie algebra $su(2)$ in this one.  
We leave this case to the readers.

By the way we are very interested in Geometric Quantum Computer, 
in particular,  Holonomic Quantum Computer which has been proposed by 
Zanardi and Rassetti \cite{ZR}, \cite{PZR} and developed by Fujii 
\cite{KF1}, \cite{KF2}, \cite{KF3} and Pachos and Chountasis \cite{PC}.   
As a general introduction to Quantum Computer (Computation) \cite{AS} 
or \cite{RP} are recommended. 

The hidden aim of this letter is to apply some results in this letter 
to Theory of Quantum Computer, in particular , quantum measurement or 
quantum entanglement or etc. See \cite{MP} and its references.

The details of this letter will be published elsewhere \cite{KF5}.

\section{Coherent and Generalized Coherent Operators}
Let $a(a^\dagger)$ be the annihilation (creation) operator of the harmonic 
oscillator.
If we set $N\equiv a^\dagger a$ (:\ number operator), then
\begin{equation}
  \label{eq:2-1}
  [N,a^\dagger]=a^\dagger\ ,\
  [N,a]=-a\ ,\
  [a^\dagger, a]=-\mathbf{1}\ .
\end{equation}
Let $\calh$ be a Fock space generated by $a$ and $a^\dagger$, and
$\{\ket{n}\vert\  n\in\futon\cup\{0\}\}$ be its basis.
The actions of $a$ and $a^\dagger$ on $\calh$ are given by
\begin{equation}
  \label{eq:2-2}
  a\ket{n} = \sqrt{n}\ket{n-1}\ ,\
  a^{\dagger}\ket{n} = \sqrt{n+1}\ket{n+1}\ ,
  N\ket{n} = n\ket{n}
\end{equation}
where $\ket{0}$ is a normalized vacuum ($a\ket{0}=0\  {\rm and}\  
\langle{0}\vert{0}\rangle = 1$). From (\ref{eq:2-2})
state $\ket{n}$ for $n \geq 1$ are given by
\begin{equation}
  \label{eq:2-3}
  \ket{n} = \frac{(a^{\dagger})^{n}}{\sqrt{n!}}\ket{0}\ .
\end{equation}
These states satisfy the orthogonality and completeness conditions
\begin{equation}
  \label{eq:2-4}
   \langle{m}\vert{n}\rangle = \delta_{mn}\ ,\quad \sum_{n=0}^{\infty}
   \ket{n}\bra{n} = \mathbf{1}\ . 
\end{equation}
 
 Let us state coherent states. For the normalized state $\ket{z} \in 
\calh \ {\rm for}\  z \in \fukuso$ the following three conditions are 
equivalent :
\begin{eqnarray}
  \label{eq:2-5-1}
 &&(\mbox{i})\quad a\ket{z} =  z\ket{z}\quad {\rm and}\quad 
      \langle{z}\vert{z}\rangle = 1  \\
  \label{eq:2-5-2}
 &&(\mbox{ii})\quad  \ket{z} =  \mbox{e}^{- \vert{z}\vert^{2}/2} 
          \sum_{n=0}^{\infty}\frac{z^{n}}{\sqrt{n!}}\ket{n} = 
          \mbox{e}^{- \vert{z}\vert^{2}/2}e^{za^{\dagger}}\ket{0} \\
  \label{eq:2-5-3}
 &&(\mbox{iii})\quad  \ket{z} =  \mbox{e}^{za^{\dagger}- \bar{z}a}\ket{0}. 
\end{eqnarray}
In the process from (\ref{eq:2-5-2}) to (\ref{eq:2-5-3}) 
we use the famous elementary Baker-Campbell-Hausdorff formula
\begin{equation}
  \label{eq:2-6}
 \mbox{e}^{A+B}=\mbox{e}^{-\frac1{2}[A,B]}\mbox{e}^{A}\mbox{e}^{B}
\end{equation}
whenever $[A,[A,B]] = [B,[A,B]] = 0$, see \cite{KS}. This is the key formula.

\noindent{\bfseries Definition}\quad The state $\ket{z}$ that 
satisfies one of (i) or (ii) or (iii) above is called the coherent state.

\noindent
The important feature of coherent states is the following partition 
(resolution) of unity.
\begin{equation}
  \label{eq:2-7}
  \int_{\fukuso} \frac{[d^{2}z]}{\pi} \ket{z}\bra{z} = 
  \sum_{n=0}^{\infty} \ket{n}\bra{n} = \mathbf{1}\ ,
\end{equation}
where we have put $[d^{2}z] = d(\mbox{Re} z)d(\mbox{Im} z)$ for simplicity.

Since the operator 
\begin{equation}
  \label{eq:2-8}
      U(z) = \mbox{e}^{za^{\dagger}- \bar{z}a}
      \quad \mbox{for} \quad z \in \fukuso  
\end{equation}
is unitary, we call this a (unitary) coherent operator. For these 
operators the following properties are crucial.
For  $z,\ w \in \fukuso$
\begin{eqnarray}  
  \label{eq:2-9-1} 
 && U(z)U(w) = \mbox{e}^{z\bar{w}-\bar{z}w}\ U(w)U(z), \\
  \label{eq:2-9-2}
 && U(z+w) = \mbox{e}^{-\frac{1}{2}(z\bar{w}-\bar{z}w)}\ U(z)U(w).
\end{eqnarray}
 
In the following section we list several basic properties of this operator.

Next let us state generalized coherent states.
Let $\{ k_+, k_-, k_3 \}$ be a Weyl basis of Lie algebra $su(1,1) 
\subset sl(2,\fukuso)$, 
\begin{equation}
  \label{eq:2-10}
 k_+ = \left(
        \begin{array}{cc}
               0 & 1 \\
               0 & 0 \\
        \end{array}
       \right),
 \quad 
 k_- = \left(
        \begin{array}{cc}
               0 & 0 \\
              -1 & 0 \\
        \end{array}
       \right),
 \quad 
 k_3 = \frac12 
       \left(
        \begin{array}{cc}
               1 &  0  \\
               0 & -1 \\
        \end{array}
       \right). 
\end{equation}
Then we have 
\begin{equation}
  \label{eq:2-11}
 [k_3, k_+]=k_+, \quad [k_3, k_-]=-k_-, \quad [k_+, k_-]=-2k_3.
\end{equation}
We note that $(k_+)^{\dagger}=-k_-$.

Next we consider a spin $K\ (> 0)$ representation of $su(1,1) 
\subset sl(2,\fukuso)$ and set its generators 
$\{ K_+, K_-, K_3 \}\ ((K_+)^{\dagger} = K_-\ \mbox{in this case})$, 
\begin{equation}
  \label{eq:2-12}
 [K_3, K_+]=K_+, \quad [K_3, K_-]=-K_-, \quad [K_+, K_-]=-2K_3.
\end{equation}
We note that this (unitary) representation is necessarily infinite 
dimensional. 
The Fock space on which $\{ K_+, K_-, K_3 \}$ act is 
$\calh_K \equiv \{\kett{K}{n} \vert n\in\futon\cup\{0\} \}$ and 
whose actions are
\begin{eqnarray}
  \label{eq:2-13}
 K_+ \kett{K}{n} &=& \sqrt{(n+1)(2K+n)}\kett{K}{n+1} , \nonumber \\
 K_- \kett{K}{n} &=& \sqrt{n(2K+n-1)}\kett{K}{n-1} ,    \\
 K_3 \kett{K}{n} &=& (K+n)\kett{K}{n}, \nonumber
\end{eqnarray}
where $\kett{K}{0}$ is a normalized vacuum ($K_-\kett{K}{0}=0$ and 
$\langle K,0|K,0 \rangle =1$). We have written $\kett{K}{0}$ instead 
of $\ket{0}$  to emphasize the spin $K$ representation, see \cite{FKSF1}. 
From (\ref{eq:2-11}), states $\kett{K}{n}$ are given by 
\begin{equation}
  \label{eq:2-14}
 \kett{K}{n} =\frac{(K_+)^n}{\sqrt{n!(2K)_n}}\kett{K}{0} ,
\end{equation}
where $(a)_n$ is the Pochammer's notation
\begin{equation}
  \label{eq:2-15}
 (a)_n \equiv  a(a+1) \cdots (a+n-1).
\end{equation}
These states satisfy the orthogonality and completeness conditions 
\begin{equation}
  \label{eq:2-16}
  \langle K,m \vert K,n \rangle =\delta_{mn}, 
 \quad \sum_{n=0}^{\infty}\kett{K}{n}\braa{K}{n}\ = \mathbf{1}_K.
\end{equation}
Now let us consider a generalized version of coherent states : 

\noindent{\bfseries Definition}\quad We call a state  
\begin{equation}
   \label{eq:2-17}
 \ket{z}  \equiv \mbox{e}^{zK_+ - \bar{z}K_-} \kett{K}{0}  
  \quad \mbox{for} \quad z \in \fukuso.
\end{equation}
the generalized coherent state (or the coherent state of Perelomov's 
type based on $su(1,1)$ in our terminology).

\noindent  
This is the extension of (\ref{eq:2-5-3}).  See the book \cite{AP}.

Then the partition of unity corresponding to (\ref{eq:2-7}) is 
\begin{eqnarray}
  \label{eq:2-18}  
    &&\int_{\fukuso} \frac{2K-1}{\pi} \frac{\mbox{tanh}(\zetta)[d^{2}z]}
     {\left(1-\mbox{tanh}^2(\zetta)\right)\zetta}
     \ket{z}\bra{z}  \nonumber \\
   = &&\int_{\mbox{D}}\frac{2K-1}{\pi} \frac{[d^{2}\zeta]}{\left(1- \vert
    \zeta\vert^{2}\right)^2} \ket{\zeta}\bra{\zeta} = 
    \sum_{n=0}^{\infty}\kett{K}{n}\braa{K}{n}\ = \mathbf{1}_K,
\end{eqnarray}
where 
\begin{equation}
  \label{eq:2-19}  
    \fukuso \rightarrow \mbox{D} : z \mapsto \zeta = \zeta(z) \equiv 
    \frac{\mbox{tanh}(\zetta)}{\zetta}z\quad
 \mbox{and}\quad \ket{\zeta} \equiv \left(1 - \vert\zeta\vert^{2}\right)^{K} 
    \mbox{e}^{\zeta K_+}\kett{K}{0}.
\end{equation}
In the process of the proof we use the disentangling formula :
\begin{equation}
  \label{eq:2-20}
    \mbox{e}^{zK_{+} -\bar{z}K_{-}}  
  = \mbox{e}^{\zeta K_+}\mbox{e}^{\log (1-\vert\zeta\vert^2)K_3}
    \mbox{e}^{-\bar{\zeta}K_-}
  = \mbox{e}^{-\bar{\zeta}K_-}\mbox{e}^{-\log (1-\vert\zeta\vert^2)K_3}
    \mbox{e}^{\zeta K_+}.
\end{equation}
This is also the key formula for generalized coherent operators.  
See \cite{AP} or \cite{FS}.

Here let us construct the spin $K$--representations making 
use of Schwinger's boson method.
 
First we assign
\begin{equation}
  \label{eq:2-21}
  K_+\equiv{1\over2}\left( a^{\dagger}\right)^2\ ,\
  K_-\equiv{1\over2}a^2\ ,\
  K_3\equiv{1\over2}\left( a^{\dagger}a+{1\over2}\right)\ .
\end{equation}
Then we have
\begin{equation}
  \label{eq:2-22}
  [K_3,K_+]=K_+\ ,\
  [K_3,K_-]=-K_-\ ,\
  [K_+,K_-]=-2K_3\ .
\end{equation}
That is, the set $\{K_+,K_-,K_3\}$ gives a unitary representation of $su(1,1)$
with spin $K = 1/4\ \mbox{and}\ 3/4$, \cite{AP}. Now we also call an operator 
\begin{equation}
  \label{eq:2-23}
   S(z) = \mbox{e}^{\frac{1}{2}\{z(a^{\dagger})^2 - \bar{z}a^2\}}
   \quad \mbox{for} \quad z \in \fukuso 
\end{equation}
the squeezed operator, see the papers in \cite{KS} or the book \cite{AP}.

Next we consider the system of two-harmonic oscillators. If we set
\begin{equation}
  \label{eq:2-24}
  a_1 = a \otimes 1,\  {a_1}^{\dagger} = a^{\dagger} \otimes 1;\ 
  a_2 = 1 \otimes a,\  {a_2}^{\dagger} = 1 \otimes a^{\dagger},
\end{equation}
then it is easy to see 
\begin{equation}
  \label{eq:2-25}
 [a_i, a_j] = [{a_i}^{\dagger}, {a_j}^{\dagger}] = 0,\ 
 [a_i, {a_j}^{\dagger}] = \delta_{ij}, \quad i, j = 1, 2. 
\end{equation}
We also denote by $N_{i} = {a_i}^{\dagger}a_i$ number operators.

Now we can construct representation of Lie algebras $su(2)$ and $su(1,1)$ 
making use of Schwinger's boson method, see \cite{FKSF1}, \cite{FKSF2}. 
Namely if we set 
\begin{eqnarray}
  \label{eq:2-26-1}
  su(2) &:&\quad
     J_+ = {a_1}^{\dagger}a_2,\ J_- = {a_2}^{\dagger}a_1,\ 
     J_3 = {1\over2}\left({a_1}^{\dagger}a_1 - {a_2}^{\dagger}a_2\right), \\
  \label{eq:2-26-2}
  su(1,1) &:&\quad
     K_+ = {a_1}^{\dagger}{a_2}^{\dagger},\ K_- = a_2 a_1,\ 
     K_3 = {1\over2}\left({a_1}^{\dagger}a_1 + {a_2}^{\dagger}a_2  + 1\right),
\end{eqnarray}
then we have
\begin{eqnarray}
  \label{eq:2-27-1}
  su(2) &:&\quad
     [J_3, J_+] = J_+,\ [J_3, J_-] = - J_-,\ [J_+, J_-] = 2J_3, \\
  \label{eq:2-27-2}
  su(1,1) &:&\quad
     [K_3, K_+] = K_+,\ [K_3, K_-] = - K_-,\ [K_+, K_-] = -2K_3.
\end{eqnarray}

In the following we define (unitary) generalized coherent operators 
based on Lie algebras $su(2)$ and $su(1,1)$. 

\noindent{\bfseries Definition}\quad We set 
\begin{eqnarray}
  \label{eq:2-28-1}
  su(2) &:&\quad 
W(z) = e^{z{a_1}^{\dagger}a_2 - \bar{z}{a_2}^{\dagger}a_1}\quad 
  {\rm for}\  z \in \fukuso , \\
  \label{eq:2-28-2}
  su(1,1) &:&\quad 
V(z) = e^{z{a_1}^{\dagger}{a_2}^{\dagger} - \bar{z}a_2 a_1}\quad 
  {\rm for}\  z \in \fukuso.
\end{eqnarray}
For the details of $W(z)$ and $ V(z)$ see \cite{AP} and \cite{FKSF1}, 
or \cite{KF2} and \cite{PC}.

In the section 4 we study the basic properties (corresponding to 
those of coherent operators in section 3) of the generalized 
coherent operators.

\noindent 
Before ending this section let us ask a question.

What is a relation between (\ref{eq:2-28-2}) and (\ref{eq:2-23}) 
of generalized coherent operators based on $su(1.1)$ ?

\noindent
The answer is given by Paris \cite{MP}:

\noindent{\bfseries The Formula}\quad We have 
\begin{equation}
  \label{eq:2-29}
  W(-\frac{\pi}{4})S_{1}(z)S_{2}(-z) W(-\frac{\pi}{4})^{-1} = V(z),
\end{equation} 
where $S_{j}(z)=(\ref{eq:2-23})$ with $a_{j}$ instead of $a$. 

\noindent 
Namely, $V(z)$ is given by ``rotating'' the product $S_{1}(z)S_{2}(-z)$ 
by $ W(-\frac{\pi}{4})$. This is an interesting relation.

\section{Basic Properties of Coherent Operators}

We make a brief review of some basic properties of coherent operators  
(\ref{eq:2-8}). For the elegant proofs see the book \cite{AP}, or the 
paper \cite{GLM} and its references.

\vspace{1cm}
\noindent{\bfseries The Matrix Elements}\quad The matrix elements of 
$U(z)$  are :
\begin{eqnarray}
   \label{eq:3-1-1}
 &&(\mbox{i})\quad n \le m \quad 
   \bra{n}U(z)\ket{m} = \mbox{e}^{-\frac{1}{2}\zetta^2}\sqrt{\frac{n!}{m!}}
                 (-\bar{z})^{m-n}{L_n}^{(m-n)}(\zetta^2), \\
   \label{eq:3-1-2}
 &&(\mbox{ii})\quad n \geq m \quad 
   \bra{n}U(z)\ket{m} = \mbox{e}^{-\frac{1}{2}\zetta^2}\sqrt{\frac{m!}{n!}}
                 z^{n-m}{L_m}^{(n-m)}(\zetta^2),
\end{eqnarray}
where ${L_n}^{(\alpha)}$ is the associated Laguerre's polynomial defined by 
\begin{equation}
   \label{eq:3-2}
 {L_n}^{(\alpha)}(x)=\sum_{j=0}^{n}(-1)^j {{n+\alpha}\choose{n-j}}
                  \frac{x^j}{j!}. 
\end{equation}
In particular ${L_n}^{(0)}$ is the usual Laguerre's polynomial and these 
are related to diagonal elements of $U(z)$. 
Here let us list the generating function and orthogonality condition of 
associated Laguerre's polynomials :
\begin{eqnarray}
   \label{eq:3-3-1}
&& \frac{\mbox{e}^{-xt/(1-t)}}{(1-t)^{\alpha +1}} = \sum_{j=0}^{\infty}
   {L_n}^{(\alpha)}(x)t^{\alpha} \quad \mbox{for}\quad 
    \vert t\vert < 1,  \\
   \label{eq:3-3-2}
&& \int_{0}^{\infty}\mbox{e}^{-x}x^{\alpha}{L_n}^{(\alpha)}(x)
   {L_m}^{(\alpha)}(x) dx = \frac{\Gamma(\alpha +n+1)}{n!}\delta_{nm} \quad
    \mbox{for}\quad \mbox{Re}(\alpha) > -1.
\end{eqnarray}

\vspace{1cm}
\noindent{\bfseries The Trace}\quad The Trace of $U(z)$ is 
\[
   \mbox{Tr}U(z) = \mbox{e}^{-\frac{1}{2}\zetta^2}
               \sum_{n=0}^{\infty}{L_n}^{(0)}(\zetta^2) 
\]
from (\ref{eq:3-1-1}). Then taking a limit $t \rightarrow 1$ in 
(\ref{eq:3-3-1}) from the below we can see easily 
\[
 \sum_{n=0}^{\infty}{L_n}^{(0)}(x)= \left\{
            \begin{array}{rl}
              0      & \quad \mbox{if}\ \  x \ne 0   \\
              \infty & \quad \mbox{if}\ \  x = 0 
            \end{array} \right.
\]
From this we can guess Trace $U(z)$ = some $\delta$ function. In fact 
we have 
\begin{equation}
   \label{eq:3-4}
    \mbox{Tr}U(z) = \pi{\delta^2}(z) \equiv \pi\delta(x)\delta(y) \quad 
    \mbox{if}\  z=x+iy.
\end{equation}
For the proof we use (\ref{eq:2-7})
\begin{equation}
   \label{eq:3-5}
  \mbox{Tr}U(z) = \mbox{Tr}(U(z)\mathbf{1}) = 
  \int_{\fukuso} \frac{[d^{2}w]}{\pi} \bra{w}U(z)\ket{w}
  = \int_{\fukuso} \frac{[d^{2}w]}{\pi}\bra{0}U(w)^{-1}U(z)U(w)\ket{0}. 
\end{equation}
It is easy to calculate the right hand side of (\ref{eq:3-5}) making use of  
(\ref{eq:2-9-2}) to get (\ref{eq:3-4}).

\vspace{1cm}
\noindent{\bfseries The Glauber's Formula}\quad The typical feature 
of coherent operators is the following Glauber's formula :\quad 
Let $A$ be any observable. Then we have 

\begin{equation}
   \label{eq:3-6}
   A = \int_{\fukuso}\frac{[d^{2}z]}{\pi}\mbox{Tr}[AU^{\dagger}(z)]U(z)
\end{equation}
This formula plays an important role in the field of homodyne tomography, 
see \cite{GLM}.

\section{Basic Properties of Generalized Coherent Operators}

We in this section study the properties of generalized coherent 
operators (\ref{eq:2-28-2}) corresponding to ones of coherent 
operators. The some results may be known, but the author could not 
find such references in spite of his efforts. At any rate 
let us list our results.

\noindent In this section we take

\noindent{\bfseries The Assumption}\quad we assume that 
\begin{equation}
   \label{eq:4-1}
    2K \in \futon 
\end{equation}
because we use a differential one of spin $K$ representation of the group 
$SU(1,1)$ as the representation in (\ref{eq:2-12}). 

\vspace{1cm}
\noindent{\bfseries The Matrix Elements}\quad The matrix elements of 
$V(z)$ are :
\begin{eqnarray}
   \label{eq:4-2-1}
  (\mbox{i})\quad n \le m \quad 
   && \braa{K}{n}V(z)\kett{K}{m}= 
    \sqrt{\frac{n!m!}{(2K)_n(2K)_m}}
    \frac{(-\bar{\kappa})^{m-n}}{(1+\kappazetta^2)^{K+\frac{n+m}{2}}}
    \ \times   \nonumber \\
   &&\sum_{j=0}^{n}(-1)^{n-j}\frac{(2K-1+m-j+n-j+j)!}{(2K-1)!(m-j)!
     (n-j)!j!}(1+\kappazetta^2)^j(\kappazetta^2)^{n-j}, \\ 
   \label{eq:4-2-2}
  (\mbox{ii})\quad n \geq m \quad    
    && \braa{K}{n}V(z)\kett{K}{m}=
    \sqrt{\frac{n!m!}{(2K)_n(2K)_m}}
    \frac{\kappa^{n-m}}{(1+\kappazetta^2)^{K+\frac{n+m}{2}}}
    \ \times   \nonumber \\
   &&\sum_{j=0}^{m}(-1)^{m-j}\frac{(2K-1+m-j+n-j+j)!}{(2K-1)!(m-j)!
     (n-j)!j!}(1+\kappazetta^2)^j(\kappazetta^2)^{m-j}, 
\end{eqnarray}
where 
\begin{equation}
   \label{eq:4-3}
\kappa \equiv \frac{\mbox{sinh}(\zetta)}{\zetta}z 
       ={\mbox{cosh}(\zetta)}\zeta.
\end{equation}

A comment is in order. The author doesn't know whether or not the right  
hand sides of (\ref{eq:4-2-1}) and (\ref{eq:4-2-2}) could be written 
making use of some special functions such as generalized Laguerre's 
functions in (\ref{eq:3-1-1}) and (\ref{eq:3-1-2}).

\vspace{1cm}
\noindent{\bfseries The Trace}\quad The Trace of $V(z)$ is 
\[
   \mbox{Tr}V(z) = 
  \sum_{n=0}^{\infty}\left\{\frac{\frac{n!}{(2K)_n}}{(1+\kappazetta^2)^{K+n}}
  \sum_{j=0}^{n}(-1)^{n-j}\frac{(2K-1+2(n-j)+j)!}{(2K-1)!{(n-j)!}^2
     j!}(1+\kappazetta^2)^j(\kappazetta^2)^{n-j} \right\}
\]
from (\ref{eq:4-2-1}) or (\ref{eq:4-2-2}), but it seems not easy to get 
a compact form (at least to the author). As an another method we use 
(\ref{eq:3-5}). From (\ref{eq:2-17}) and (\ref{eq:2-18}) we have 
\begin{eqnarray}
   \label{eq:4-4}
  \mbox{Tr}V(z) &=& \mbox{Tr}(V(z)\mathbf{1_K}) =  
   \int_{\fukuso} \frac{2K-1}{\pi} \frac{\mbox{tanh}(\wzetta)[d^{2}w]}
     {\left(1-\mbox{tanh}^2(\wzetta)\right)\wzetta}
    \bra{w}V(z)\ket{w} \nonumber \\
  &=& 
   \int_{\fukuso} \frac{2K-1}{\pi} \frac{\mbox{tanh}(\wzetta)[d^{2}w]}
     {\left(1-\mbox{tanh}^2(\wzetta)\right)\wzetta}
    \braa{K}{0}V(w)^{-1}V(z)V(w)\kett{K}{0}. 
   \end{eqnarray}

We can perform the calculation of (\ref{eq:4-4}) in spite of very 
hard one.

\noindent
Making use of change of variables $w \mapsto \zeta = \zeta(w) \equiv 
\frac{\mbox{tanh}(\wzetta)}{\wzetta}w$ in (\ref{eq:2-19}) and 
disentangling formula (\ref{eq:2-20}) we have
\begin{equation}
   \label{eq:4-5}
\mbox{Tr}V(z) 
  = \frac{1}{\left\{\mbox{cosh}(\zetta)\right\}^{2K}} 
   \int_{\mbox{D}}\frac{2K-1}{\pi} \frac{[d^{2}\zeta]}{\left(1- \vert
    \zeta\vert^{2}\right)^2} \frac{1}{ \left\{ 1 
    - \frac{\mbox{tanh}(\zetta)}{\zetta} \frac{\bar{z}\zeta-z\bar{\zeta}}
    {1-\vert\zeta\vert^2} \right\}^{2K} }.
\end{equation}
Moreover we set $\chi \equiv \frac{\mbox{tanh}(\zetta)}{\zetta}z$ 
for simplicity, then we have 
\begin{equation}
   \label{eq:4-6}
\mbox{Tr}V(z) =
  (1- \chizeta^2)^{K}
   \int_{\mbox{D}}\frac{2K-1}{\pi} \frac{[d^{2}\zeta]}{\left(1- \vert
    \zeta\vert^{2}\right)^2} \frac{1}{ \left( 1 - 
 \frac{\bar{\chi}\zeta-\chi\bar{\zeta}}{1-\vert\zeta\vert^2} \right)^{2K} }.
\end{equation}
To calculate this we moreover assume  $2K \in \futon - \{1\}$ 
because the measure of the integral contains the term $2K-1=0$ !. 

\noindent Under this assumption we can perform the integral and obtain 
\begin{equation}
   \label{eq:4-7}
   \int_{\mbox{D}}\frac{2K-1}{\pi} \frac{[d^{2}\zeta]}{\left(1- \vert
    \zeta\vert^{2}\right)^2} \frac{1}{ \left( 1 - 
 \frac{\bar{\chi}\zeta-\chi\bar{\zeta}}{1-\vert\zeta\vert^2} \right)^{2K} }
   = \frac{1}{2\chizeta(1+\chizeta)^{2K-1}}.
\end{equation}

\noindent
As a result we have 
\begin{equation}
   \label{eq:4-8}
\mbox{Tr}V(z) 
        = \frac{(1-\chizeta^2)^{K}}{2\chizeta(1+\chizeta)^{2K-1}}
        = \frac{1+\chizeta}{2\chizeta}\left(\frac{1-\chizeta}
                {1+\chizeta}\right)^{K}.
\end{equation}

A comment is in order. The author believes that 

\noindent{\bfseries Conjecture}\quad we have for all\ \ $2K\geq 1$ 
\begin{equation}
   \label{eq:4-9}
   \int_{\mbox{D}}\frac{2K-1}{\pi} \frac{[d^{2}\zeta]}{\left(1- \vert
    \zeta\vert^{2}\right)^2} \frac{1}{ \left( 1 - 
 \frac{\bar{\chi}\zeta-\chi\bar{\zeta}}{1-\vert\zeta\vert^2} \right)^{2K} }
  = \frac{1}{2\chizeta(1+\chizeta)^{2K-1}}.
\end{equation}

\noindent
Namely we believe that the formula (\ref{eq:4-8}) holds for all $2K \geq 1$.

\vspace{1cm}
\noindent{\bfseries The Glauber's Formula}\quad Unfortunately in this 
case we have no Glauber's formula : For a observable $A$ 

\begin{equation}
   \label{eq:4-10}
   A \ne \int_{\fukuso}\frac{2K-1}{\pi}\frac{\mbox{tanh}(\zetta)[d^{2}z]}
     {\left(1-\mbox{tanh}^2(\zetta)\right)\zetta}
     \mbox{Tr}[AV^{\dagger}(z)]V(z)
\end{equation}

\noindent 
In fact if we set $A = \kett{K}{0}\braa{K}{0}$ then it is not difficult 
to check (\ref{eq:4-10}) making use of (\ref{eq:4-2-1}) and (\ref{eq:4-2-2}).

\section{Discussion}

In this paper we listed the three basic properties of coherent operators 
and investigated these ones for generalized coherent operators based on 
$su(1,1)$ $\cdots$ $V(z)$ in (\ref{eq:2-28-1}). 
But we have not investigated these for generalized coherent operators 
based on $su(2)$ $\cdots$ $W(z)$ in (\ref{eq:2-28-2}). 
We leave these to the readers. 

The several caluculations performed in Section 4 are not easy. We will 
publish these ones in \cite{KF5}.

\noindent{\em Acknowledgment.}\\
The author wishes to thank K. Funahashi for his helpful comments and 
suggestions.

\vspace{10mm}

\noindent{{\bfseries [Note added in proof]}}
\vspace{2mm}

From (\ref{eq:4-8}) with $\chizeta = \mbox{tanh}(\zetta)$ it is easy 
to see 
\begin{equation}
   \label{eq:4-11}
\mbox{The right hand side of (\ref{eq:4-8})} 
     =\frac{\mbox{e}^{-2\zetta K}}{1-\mbox{e}^{-2\zetta}}
     =\sum_{n=0}^{\infty}\mbox{e}^{-2\zetta(K+n)}
\end{equation}
, so that from (\ref{eq:2-13}) we can conjecture 
$\mbox{Tr}V(z)=\mbox{Tr}\ \mbox{e}^{-2\zetta K_3}$ (this was pointed out 
by S. Sakoda. The author thanks him for this suggestion.) 
What is this meaning ?  After some considerations we reached

\noindent{\bfseries Formula}\quad we have the (new ?) decomposition 
formula 
\begin{equation}
   \label{eq:4-12}
   \mbox{e}^{zK_{+} - \bar{z}K_{-}}
   =\mbox{e}^{\frac{\pi}{4} 
       \left( \frac{z}{\zetta}K_{+}+\frac{\bar{z}}{\zetta}K_{-} \right)
             }
    \mbox{e}^{-2\zetta K_{3}}
     \mbox{e}^{-\frac{\pi}{4} 
       \left( \frac{z}{\zetta}K_{+}+\frac{\bar{z}}{\zetta}K_{-} \right)
              }
\end{equation}     
for all $2K \geq 1$ (!).

\noindent A comment is in order. As far as the author knows this 
decomposition formula has not been used in the references. 

\noindent Therefore we obtain
\begin{equation}
   \label{eq:4-13}
    \mbox{Tr}V(z)=\mbox{Tr}\ \mbox{e}^{-2\zetta K_3}
                 =\mbox{The right hand side of (\ref{eq:4-8})} 
\end{equation}  
for all $2K \geq 1$. Namely we could solve our conjecture (\ref{eq:4-9}) 
indirectly. But the direct proof is still unknown. 

\noindent{\bfseries Problem}\quad Prove directly for all\ \ $2K\geq 1$ 
\[
   \int_{\mbox{D}}\frac{2K-1}{\pi} \frac{[d^{2}\zeta]}{\left(1- \vert
    \zeta\vert^{2}\right)^2} \frac{1}{ \left( 1 - 
 \frac{\bar{\chi}\zeta-\chi\bar{\zeta}}{1-\vert\zeta\vert^2} \right)^{2K} }
  = \frac{1}{2\chizeta(1+\chizeta)^{2K-1}}. 
\]
We note that in the case $2K \in \futon - \{1\}$ we have used the residue 
theorem in Complex Analysis.


\end{document}